\documentclass{article}
\usepackage[margin=2cm]{geometry}
\pdfoutput=1 

\usepackage{microtype}
\date{}

\usepackage{amsmath}
\usepackage{amsfonts}
\usepackage{amsthm}
\usepackage{booktabs}
\usepackage{xspace}
\usepackage{subfig}

\usepackage{graphicx}
\usepackage[tiny,compact]{titlesec}
\usepackage{url}
\usepackage{paralist}
\usepackage{enumitem}
\usepackage{natbib}
\usepackage{color}
\usepackage{hyperref}

\usepackage{adjustbox}
\usepackage{fancyvrb}
\newenvironment{myverb}{%
 \VerbatimEnvironment
 \begin{adjustbox}{max width=\linewidth}
 \begin{BVerbatim}
  }{
  \end{BVerbatim}
 \end{adjustbox}
}

\title{Citation Count Analysis for Papers with Preprints}
\author{Sergey Feldman, Kyle Lo, Waleed Ammar \\
{\normalsize Allen Institute of Artificial Intelligence, Seattle, WA} \\
{\normalsize \url{waleeda@allenai.org}}}

\begin{document}
\maketitle

\section*{Abstract}
We explore the degree to which papers prepublished on arXiv garner more citations, in an attempt to paint a sharper picture of fairness issues related to prepublishing.
A paper's citation count is estimated using a negative-binomial generalized linear model (GLM) while observing a binary variable which indicates whether the paper has been prepublished.
We control for author influence (via the authors' h-index at the time of paper writing), publication venue, and overall time that paper has been available on arXiv.
Our analysis only includes papers that were eventually accepted for publication at top-tier CS conferences, and were posted on arXiv either before or after the acceptance notification.
We observe that papers submitted to arXiv before acceptance have, on average, 65\% more citations in the following year compared to papers submitted after. We note that this finding is not causal, and discuss possible next steps.


\section{Introduction}\label{introduction}
Preprint servers like arXiv enable researchers to self-distribute scientific paper drafts with minimal moderation.
While some of these papers are never published elsewhere, many are also accepted for publication at academic venues after a double-blind peer-review process.
Authors of these papers are faced with the decision to distribute their papers on arXiv \textit{before} or \textit{after} acceptance at their target publication venues. We refer to those papers that are posted on arXiv before acceptance as \emph{prepublished}.

With the increasing popularity of prepublishing computer science (CS) papers on arXiv \citep{Sutton2017Popularity}, this decision has been the subject of a considerable debate in the CS research community (among others).\footnote{See Marti Hearst's and Kelly Cruz's thoughtful discussions of this topic at \url{https://acl2017.wordpress.com/2017/02/19/arxiv-and-the-future-of-double-blind-conference-reviewing/} and {http://www.astrobetter.com/blog/2011/12/12/to-post-or-not-to-post-publishing-to-the-arxiv-before-acceptance/}.}
Some researchers abstain from posting their work on arXiv until it has been accepted for publication at the target venue.  One reason is to preserve author anonymity during the double-blind review process and mitigate reviewer bias favoring well-known authors and affiliations \citep[e.g.,][]{snodgrass2006,tomkins2017}. Other reasons may include fear of circulating incorrect results or conclusions or fear of retaliation by a reviewer in conflict with the author.

On the other side of the debate, some researchers prefer publishing drafts of their work on arXiv before it is accepted for publication.  One reason is to allow other researchers to build on their work which can expedite scientific developments.
Another reason is to allow researchers aside from the official reviewers at the target venue to provide feedback, which can be used to further improve the paper even before it is published (i.e., before the camera-ready due date).
Authors also may use arXiv for ``flag-planting'', i.e., claiming a research contribution before getting scooped by other researchers who may be doing similar work.

Quoting a recent blog post by Yoav Goldberg: \textit{``[T]here is also a rising trend of people using arXiv for flag-planting, and to circumvent the peer-review process. This is especially true for work coming from `strong' groups. Currently, there is practically no downside to posting your (often very preliminary, often incomplete) work to arXiv, only potential benefits.''}\footnote{\url{https://medium.com/@yoav.goldberg/an-adversarial-review-of-adversarial-generation-of-natural-language-409ac3378bd7}}
In this work, our goal is to quantify some of these perceived benefits of posting a paper on arXiv before it is submitted for publication.
While it may be hard to study reviewer bias for prepublished papers (since the review results are not made public), we can observe the number of times a paper is cited, which is often used to measure a paper's impact.\footnote{While most peer reviews are not publicly available, a notable exception is the International Conference on Learning Representations (ICLR) which makes all reviews available and also allows any researcher to comment on papers under submission using the \url{openreview.net} platform.}
We focus on papers with an arXiv-published draft which have also been accepted for publication at a top-tier venue.  Specifically, we're interested in studying whether there are significant differences in citation counts between papers that were published on arXiv before~vs.~after they were accepted at the venue at which they were eventually published. 

To motivate this study, consider the following scenario: Two researchers R1 and R2 worked independently developed an outstanding method around the same time. 
R1 decides to prepublished her draft on arXiv while R2 decides to wait until the paper is accepted for publication at target venue. 
Naturally, the earlier exposure of the research community to R1's paper may result in researchers attributing most of the credit to R1 rather than R2, despite both being eventually published at the same venue. 
We may consequently observe a higher number of citations for R1's work rather than R2's work. 
This is especially concerning when metrics derived from citation counts (e.g. h-index) play a significant role in hiring and promotion decisions in universities and research labs (despite the controversy surrounding number of citations as a measure of a paper's impact).

In this draft, we explore the degree to which prepublished papers garner more citations, in an attempt to paint a sharper picture of arXiv-related fairness issues.
We use a negative-binomial generalized linear model (GLM) to regress a paper's number of citations onto a binary indicator representing arXiv prepublishing, and control for author influence (via the authors' h-index at the time of paper writing), publication venue, and overall time that paper has been available on arXiv.
We analyze papers that were eventually accepted for publication at top-tier CS conferences, and were posted on arXiv either before or after the acceptance notification.  
We observe a significant positive association between citation count and prepublishing on arXiv.

Our results are consistent with previous work \cite[e.g.,][]{larivire:14} which found papers posted on arXiv to have higher citation rate (among all papers published in Web of Science).\footnote{\url{http://webofknowledge.com/}}
Also related is \citet{moed:07} who  studied the higher citation rate of arXiv papers and found a strong quality bias and early view effect, and found no effect due to the open access nature of arXiv.
To the best of our knowledge, this is the first study to analyze the pre-publication effect, distinguishing between papers posted on arXiv before vs.~after conference acceptance.
We also control for important aspects such as author popularity and venues which are known to affect citation rates.

\section{Data}\label{data}
Here, we describe the data we used for this analysis in some detail.

\subsection{Venues}
All papers included in our study were eventually published at one of the following top-tier computer science conferences, which have a significant portion of their papers on arXiv:  AAAI, ACL, CVPR, ECCV, EMNLP, FOCS, HLT-NAACL, ICCV, ICML, ICRA, IJCAI, INFOCOM, KDD, NIPS, SODA and WWW.  
We include papers published since 2007 and no later than 2016, so that we can count the number of citations they receive during the year following their publication.

To obtain this data, we queried Semantic Scholar for all the papers published in a particular conference. We then looked up each of these papers in the arXiv metadata dump contributed by \citet{Sutton2017Popularity},\footnote{https://github.com/casutton/cs-arxiv-popularity-code} and obtained arXiv submission dates for each paper that was posted.
For papers with multiple versions on arXiv, we record the date of the earliest submission, and papers that were never posted to arXiv were excluded.

See Table \ref{tab:conf_numbers} for a per-conference break down of the 4392 papers in our dataset.

\begin{table}[h!]
\centering
\begin{tabular}{ l | c  }
Venue & No. of Papers \\
\hline
AAAI      &     3726 \\
NIPS      &     3393 \\
IJCAI     &     3001 \\ 
WWW       &     2958 \\
ACL       &     2676 \\
ICML      &     2200 \\
KDD       &     1661 \\
ECCV      &     1477 \\
EMNLP     &     1248 \\
SODA      &     1234 \\
HLT-NAACL &     876 \\
CVPR      &     467 \\
FOCS      &     305 \\
INFOCOM   &     183 \\ 
ICRA      &     182 \\
ICCV      &     156
\end{tabular}
\caption{Number of papers from each conference in our dataset.}
\label{tab:conf_numbers}
\end{table}

We used the Calls for Papers Wiki \footnote{\url{http://wikicfp.com}} to obtain paper submission deadlines for most of the conference and year combinations in our dataset. The rest were obtained via web search. 

\subsection{Citations}
\begin{figure}[!tbp]
  \centering
  \subfloat[Histogram of the number of paper citations in the year following a conference.]{\includegraphics[width=0.47\textwidth]{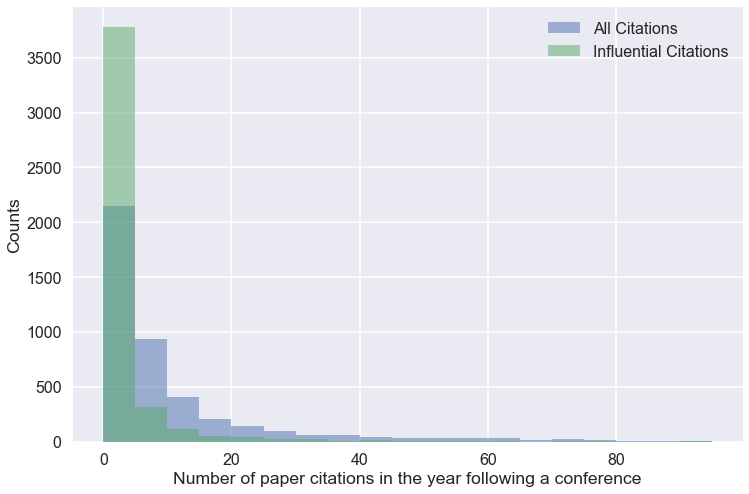}\label{fig:citations_hist}}
  \hfill
  \subfloat[Histogram of the difference between the arXiv submission date and the conference deadline.]{\includegraphics[width=0.47\textwidth]{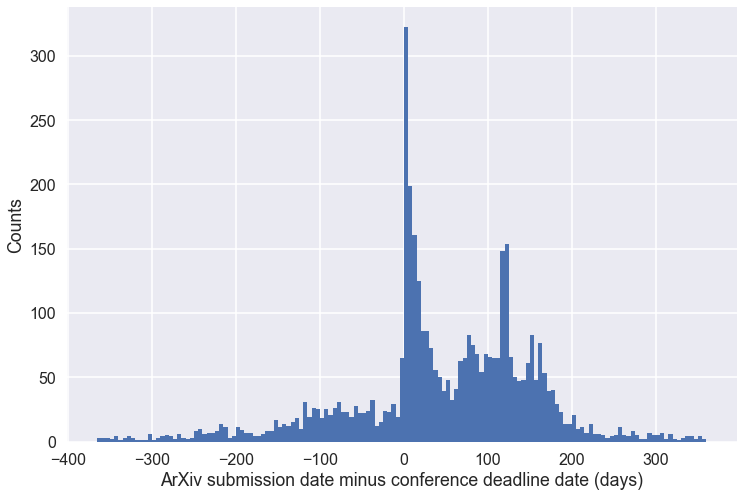}\label{fig:sub_minus_dl_hist}}
\end{figure}

The response variable we would like to model is the number of times a paper is cited in the calendar year following the conference, which we label as ``all citations.''\footnote{Alternatively, we could have simply counted all citations a paper received but this would require making stronger assumptions about how the number of citations change over years, which is not the focus of this study.}
Figure \ref{fig:citations_hist} shows the histogram of citation count with buckets of size 5, showing that the vast majority of papers in this population receive fewer than 20 citations in the calendar year following the conference.

We also experiment with a modified definition of the response variable meant to count meaningful citations (e.g., omitting self citations), which we label as ``Influential Citations'' to distinguish it from ``All Citations.'' Our definition of influential citations is based on \citet{hics}, and only counts citations with no overlap in the author lists. In an influential citation, the cited paper is referenced three times or more in the narrative of the citing paper, not consistently combined with other references, mentioned in context of experimental results, or explicitly mentioned as foundation for the citing paper.

\subsection{Author influence}
We suspect that well known authors tend to garner more citations than less known authors. 
In order to control for this source of bias in our analysis, we model an observed variable which represents the author’s influence. Given the paper in question, we first compute the h-index for its authors one year before it was published. Then we take the maximum h-index among all the authors of a paper and use this single value as a per-paper summary for author influence. 
Let $h(a, \text{year})$ be the h-index for author $a$ at a specified year. The author influence for paper $p$ can then be written as:

$$h_\text{max}(p) = \max_{a \in \text{authors}(p)} h(a, \text{year}(p))$$

Because h-index is non-linear in its relationship with citation counts, we model it as a categorical variable with ten buckets each of which containing the same number of papers.
The first bucket included all papers with $h_\text{max}(p) \leq 6$ and the last bucket included all papers with $h_\text{max}(p) \geq 42$.

\subsection{Time available on arXiv}
Papers prepublished on arXiv before acceptance have had more time to gather citations than those posted to arXiv after acceptance, which may explain any differences in citation counts.  To control for this factor, we compute the fraction of the year the paper has been available on arXiv.
In particular, we measure the number of days between the first arXiv submission and the beginning of the calendar year in which we count citations of that paper, then divide by the number of days in the year, as illustrated by the following Python code.

\begin{verbatim}
next_year_jan_1 = datetime(year=conf_year + 1, month=1, day=1).date()
delta = next_year_jan_1 - arxiv_submission_date
frac_year_remaining = np.maximum(delta.days / 365, 0)
\end{verbatim}

We clamp the difference (\verb|delta.days|) at a minimum of zero because a paper may be put on arXiv for the first time long after it is officially published.


\subsection{Submitted to arXiv before vs.~after acceptance}
This variable is an indicator for whether the paper was posted to arXiv before or after it was accepted for publication.
Ideally, we'd like to observe whether the arXiv submission date is before or after the acceptance notification, but since the exact acceptance dates were not available for all venues, we use a conservative estimate of +28 days after the the submission deadline of the conference as our prepublishing threshold.  
Figure \ref{fig:sub_minus_dl_hist} contains a histogram showing the distribution of arXiv submission dates relative to the paper's target venue deadline date.


\subsection{Summary of variables}

To summarize, we compute the following variables for each paper $p$:
 
\begin{itemize}
\item{\verb|cites_1year|} - number of papers that cited $p$ and were published in the calendar year following the official publication of $p$ (continuous response variable).
\item{\verb|influential_cites_1year|} - number of papers that cited $p$ and were published in the calendar year following the official publication of $p$ and satisfied `influential' criteria (continuous response variable).
\item{\verb|max_hindex_decile|} - the decile into which the maximum (across all authors) h-index of $p$ falls into (categorical feature - 10 values).
\item{\verb|submitted_before_deadline|} - whether $p$ was submitted 28 days after the conference submission deadline (binary feature).
\item{\verb|frac_year_remaining|} - fraction of year remaining from arXiv submission date until the year after the conference in which paper $p$ was published (continuous feature).
\item{\verb|conf|} - the conference where $p$ was published (categorical feature - 16 values).
\end{itemize}

\section{Analysis}\label{analysis}
Here, we describe how we model the variables discussed in the previous section then analyze the results.

\subsection{Model}
Negative binomial GLMs are a common option for modeling count-valued response variables that exhibit \emph{overdispersion} (i.e. when variance of the variable exceeds its mean, thus deviating from the standard Poisson count model) which is typical of real-world data \citep{hilbe2007}. One can interpret the negative binomial distribution as a marginalized Poisson distribution where its mean is drawn from a Gamma distribution.  

The conditional mean model is expressed as:
$$E[y|\mathbf{x}] = \exp\left(w_0 + \sum_i w_ix_i\right),$$
where $y$ is the response variable, $\mathbf{x}$ is the vector of covariates/features, and $w_i$ is the learned weight of the $i$th feature $x_i$.  In our case, the response variable $y$ is either \verb|cites_1year| or \verb|influential_cites_1year|.  Within our feature vector $\mathbf{x}$, our primary covariate of interest is \verb|submitted_before_deadline|, while the other features are possible confounders that we want to control for.  

We use Python's \verb|statsmodels| \citep{statsmodels2010} to fit the following regression models (expressed in the standard formula mini-language from \verb|R| that is also used in \verb|statsmodels|): 

\begin{myverb}

cites_1year ~ max_hindex_decile + frac_year_remaining + conf

cites_1year ~ max_hindex_decile + frac_year_remaining + conf + submitted_before_deadline

\end{myverb}

The only difference between these two models is the presence of the \verb|submitted_before_deadline| binary variable.  We repeat this again for \verb|influential_cites_1year| as the response variable.

\subsection{Results}

We conducted a likelihood ratio test on the two models and the resulting p-value was tiny: $6.27\mathrm{e}{-29}$. This means that the second model has a significantly higher likelihood, indicating that it better fits the data. The coefficients of the full model that includes \verb|submitted_before_deadline| are shown below:

\begin{myverb}

                 Generalized Linear Model Regression Results                  
==============================================================================
Dep. Variable:            cites_1year   No. Observations:                 4392
Model:                            GLM   Df Residuals:                     4365
Model Family:        NegativeBinomial   Df Model:                           26
Link Function:                    log   Scale:                   3.30268468922
Method:                          IRLS   Log-Likelihood:                -14832.
Date:                Mon, 12 Mar 2018   Deviance:                       6376.3
Time:                        11:46:30   Pearson chi2:                 1.44e+04
No. Iterations:                    11                                         
=====================================================================================================
                                        coef    std err          z      P>|z|      [0.025      0.975]
-----------------------------------------------------------------------------------------------------
Intercept                             0.9192      0.198      4.634      0.000       0.530       1.308
max_hindex_decile[T.(6.0, 10.0]]      0.2249      0.160      1.408      0.159      -0.088       0.538
max_hindex_decile[T.(10.0, 13.0]]     0.3543      0.165      2.147      0.032       0.031       0.678
max_hindex_decile[T.(13.0, 16.0]]     0.3265      0.158      2.062      0.039       0.016       0.637
max_hindex_decile[T.(16.0, 19.0]]     0.5266      0.154      3.416      0.001       0.224       0.829
max_hindex_decile[T.(19.0, 22.0]]     0.7298      0.161      4.532      0.000       0.414       1.045
max_hindex_decile[T.(22.0, 26.0]]     0.4174      0.155      2.695      0.007       0.114       0.721
max_hindex_decile[T.(26.0, 32.0]]     0.5917      0.150      3.953      0.000       0.298       0.885
max_hindex_decile[T.(32.0, 41.0]]     0.6185      0.151      4.105      0.000       0.323       0.914
max_hindex_decile[T.(41.0, 99.0]]     1.0595      0.145      7.284      0.000       0.774       1.345
submitted_before_deadline[T.True]     0.5029      0.083      6.080      0.000       0.341       0.665
conf[T.ACL]                           1.2415      0.201      6.169      0.000       0.847       1.636
conf[T.CVPR]                          1.4699      0.155      9.488      0.000       1.166       1.773
conf[T.ECCV]                          1.4585      0.190      7.658      0.000       1.085       1.832
conf[T.EMNLP]                         0.9585      0.207      4.637      0.000       0.553       1.364
conf[T.FOCS]                          0.0017      0.178      0.010      0.992      -0.347       0.350
conf[T.HLT-NAACL]                     1.1061      0.272      4.060      0.000       0.572       1.640
conf[T.ICCV]                          1.1248      0.208      5.418      0.000       0.718       1.532
conf[T.ICML]                          0.5132      0.147      3.480      0.001       0.224       0.802
conf[T.ICRA]                         -0.0980      0.223     -0.439      0.661      -0.536       0.339
conf[T.IJCAI]                        -0.2673      0.199     -1.341      0.180      -0.658       0.123
conf[T.INFOCOM]                      -0.1444      0.202     -0.715      0.474      -0.540       0.251
conf[T.KDD]                           0.5083      0.213      2.385      0.017       0.091       0.926
conf[T.NIPS]                          0.6280      0.156      4.031      0.000       0.323       0.933
conf[T.SODA]                         -0.6441      0.165     -3.892      0.000      -0.968      -0.320
conf[T.WWW]                           0.5485      0.217      2.531      0.011       0.124       0.973
frac_year_remaining                   0.1710      0.107      1.599      0.110      -0.039       0.381
=====================================================================================================
\end{myverb}

Due to the $\exp$ term in the regression function, these coefficients can be interpreted as having a multiplicative effect instead of an additive effect as in linear regression. We can thus look at the 0.5029 coefficient of \verb|submitted_before_deadline| (the coef column), and interpret its effect as multiplying the number of citations by exp(0.5029) = 1.65. In other words, the fitted regression model estimates that papers submitted to arXiv before acceptance, on average, tend to have 65\% more citations in the following year compared to papers submitted after. 

The difference is even more pronounced when we look at the number of influential citations.\footnote{We omit detailed results for influential citations for brevity.} Papers submitted to arXiv before acceptance, on average, tend to have 75\% more influential citations in the following year compared to papers submitted after. 
We emphasize that we cannot conclude that prepublishing on arXiv has a causal effect on citation counts since this result is not based on a randomized controlled experiment.

Note that in this framework, each categorical variable with $k$ values has only $k-1$ coefficients. Each coefficient can be interpreted as being relative to some baseline value, which is determined by the left-out value.  For example, the baseline category for \verb|max_hindex_decile| is [0, 6], and the coefficients for the other nine deciles capture how many more citations one can expect to have with higher h-indices (in an associative, not causal, sense).  
In particular, an h-index between 42 and 99 is associated (on average) with more than double the number of next-year citations than if you had an h-index between 0 and 6. These coefficients increase in a nearly-monotonic way as h-index deciles increase, which is consistent with our intuition that more famous authors tend to get more citations.  
Similarly, the baseline conference is AAAI.

The results suggest that \verb|frac_year_remaining| is a minor variable, with an estimate of 0 being part of the 95\% confidence interval (last two columns).
This is somewhat surprising since we expected papers which have been on arXiv for a longer fraction of a given year to have more citations in the following year.

\section{Conclusion}\label{conclusions}
Our exploratory analysis shows that publishing a CS paper on arXiv before it is eventually accepted (as opposed to after) for publication at a top tier target venue is associated with ~65\% more citations in the calendar year following the conference. 
Although we take into account other factors which can influence number of citations (namely, author influence, publication venue, time available on arXiv), there may be other confounding factors which we did not include in our study (e.g., author affiliation, paper quality). We invite researchers interested in this analysis to explore the effect of other factors we have not included in the model, and invite conference chairs to conduct randomized controlled experiments in which authors submitting their drafts to the conference agree to prepublish their drafts on arXiv if they are randomly selected. 


We note that identifying the potential unfair advantage given to prepublished papers may not give researchers a sufficiently compelling reason to delay posting their paper drafts on arXiv until the review process has completed.
Instead, we encourage the community to adopt anonymous prepublished submissions (with pre-specified time limits on the anonymity) on arXiv and related platforms, similar to how the OpenReview platform implemented the peer reviewing process for ICLR 2018.\footnote{\url{https://iclr.cc/archive/www/doku.php\%3Fid=iclr2018:faq.html\#what_is_the_signature_field_when_submitting_a_comment_review}}


\section*{Acknowledgements}
We are grateful for the Semantic Scholar team as well as the teams behind arXiv and WikiCFP for their commitment to promoting transparency and openness in scientific communication. 
We thank Oren Etzioni, Yoav Goldberg and Mark Neumann for helpful comments.

\bibliographystyle{plainnat}
\bibliography{early_arxiv_posting}

\begin{thebibliography}{8}
\providecommand{\natexlab}[1]{#1}
\providecommand{\url}[1]{\texttt{#1}}
\expandafter\ifx\csname urlstyle\endcsname\relax
  \providecommand{\doi}[1]{doi: #1}\else
  \providecommand{\doi}{doi: \begingroup \urlstyle{rm}\Url}\fi

\bibitem[Hilbe(2007)]{hilbe2007}
Joseph~M. Hilbe.
\newblock \emph{Negative Binomial Regression}.
\newblock Cambridge University Press, 2007.
\newblock \doi{10.1017/CBO9780511811852}.

\bibitem[Larivi{\`e}re et~al.(2014)Larivi{\`e}re, Sugimoto, Macaluso,
  Milojevic, Cronin, and Thelwall]{larivire:14}
Vincent Larivi{\`e}re, Cassidy~R. Sugimoto, Benoit Macaluso, Stasa Milojevic,
  Blaise Cronin, and Mike Thelwall.
\newblock arxiv e-prints and the journal of record: An analysis of roles and
  relationships.
\newblock \emph{JASIST}, 65:\penalty0 1157--1169, 2014.

\bibitem[Moed(2007)]{moed:07}
Henk~F. Moed.
\newblock The effect of 'open access' upon citation impact: An analysis of
  arxiv's condensed matter section.
\newblock \emph{JASIST}, 58:\penalty0 2047--2054, 2007.

\bibitem[Seabold and Perktold(2010)]{statsmodels2010}
J.S. Seabold and J.~Perktold.
\newblock Statsmodels: Econometric and statistical modeling with python.
\newblock In \emph{Proceedings of the 9th Python in Science Conference}, 2010.

\bibitem[Snodgrass(2006)]{snodgrass2006}
Richard~T. Snodgrass.
\newblock Single-versus double-blind reviewing: an analysis of the literature.
\newblock \emph{SIGMOD Record}, 35:\penalty0 8--21, 2006.

\bibitem[Sutton and Gong(2017)]{Sutton2017Popularity}
Charles Sutton and Linan Gong.
\newblock Popularity of arxiv.org within computer science.
\newblock \emph{ArXiv e-prints}, abs/1710.05225, 2017.

\bibitem[Tomkins et~al.(2017)Tomkins, Zhang, and Heavlin]{tomkins2017}
Andrew Tomkins, Min Zhang, and William~D. Heavlin.
\newblock Reviewer bias in single- versus double-blind peer review.
\newblock In \emph{Proceedings of the National Academy of Sciences of the
  United States of America}, 2017.

\bibitem[Valenzuela et~al.(2015)Valenzuela, Ha, and Etzioni]{hics}
Marco Valenzuela, Vu~Ha, and Oren Etzioni.
\newblock Identifying meaningful citations.
\newblock In \emph{AAAI Workshop: Scholarly Big Data}, 2015.

\end{thebibliography}

\end{document}